\documentclass[11pt,epsfig]{amsart}
\usepackage{amsbsy,amssymb,amscd,amsfonts,latexsym,amstext,delarray,
amsmath,epsfig,graphicx}  \headsep=15pt
\input xypic

\usepackage{hyperref}

\newtheorem{thm}{Theorem}[section]
\newtheorem{prop}[thm]{Proposition}

\newtheorem{defn}[thm]{Definition}

\def\C{{\mathbb C}}

\renewcommand{\H}{{\mathbb H}}

\def\Z{{\mathbb Z}}
\def\R{{\mathbb R}}

\def\cA{{\mathcal A}}

\def\cC{{\mathcal C}}

\def\cH{{\mathcal H}}

\def\cL{{\mathcal L}}
\def\cM{{\mathcal M}}

\def\cU{{\mathcal U}}

\newcommand{\ie}{{\it i.e.\/}\ }

\newcommand{\cf}{{\it cf.\/}\ }

\def\text{\hbox}

\def\Aut{{\rm Aut}}

\def\Diff{{\rm Diff}}

\def\Inn{{\rm Inn}}

\def\Out{{\rm Out}}

\def\SU{{\rm SU}}

\def\Tr{{\rm Tr}}

\def\qqq{\,,\quad \forall\,}

\def\dirac{{\partial \hspace{-6pt} \slash}}

\def\cx{{U^{lep}}}

\def\higgs{{\bf H}}

\title
{ Noncommutative Geometry and the   standard model  with neutrino
mixing }
\author{Alain Connes }
\date{}

\begin{document}

\maketitle

\vspace{2cm}
\begin{abstract}
We show that allowing the metric dimension of a space to be
independent of its KO-dimension and turning the finite
noncommutative geometry F-- whose product with classical
4-dimensional space-time gives the standard model coupled with
gravity--into a space of KO-dimension 6 by changing the grading on
the antiparticle sector into its opposite, allows to solve three
problems of the previous noncommutative geometry interpretation of
the standard model of particle physics:
  The finite geometry F is no longer put in ``by hand" but
  a conceptual understanding of  its structure
  and a classification of its metrics is given.
  The fermion doubling problem  in
  the fermionic part of the action is resolved.
  The spectral action of our joint work with Chamseddine
  now automatically generates the full standard model coupled with gravity
  with  neutrino mixing and see-saw
  mechanism for neutrino masses. The predictions of the Weinberg angle and the
  Higgs scattering parameter at unification scale are the same as in our joint work but
  we also find a mass relation (to be imposed at unification scale).

\end{abstract}

\section{Introduction}

We showed some time ago (\cf \cite{Co-book}) how to interpret the
Lagrangian of the standard model in terms of noncommutative
geometry. This interpretation was based on the extension of the
Yang-Mills functional to the algebraic framework of NCG. In
\cite{Co-book} the color degrees of freedom were still added in an
artificial manner and the action functional was obtained by analogy
with the classical gauge theories. In our joint work with A.
Chamseddine  \cite{cc1}, \cite{cc2} and in \cite{CoSM} we showed how
to incorporate the color naturally and more importantly how to
obtain the bosonic part of the standard model action coupled to
gravity from a very general {\em spectral} action principle. We call
it a principle since it is based on the very general idea that a
refined notion of geometry (suitable in particular to deal with
spaces whose coordinates do not commute) is obtained by focussing
not on the traditional $g_{\mu\nu}$ but on the Dirac operator $D$.
As it turns out this way of defining a geometry by specifying the
Dirac operator is meaningful both in mathematical terms (where the
Dirac operator specifies the fundamental class in $KO$-homology) and
in physics terms (where, modulo a chiral gauge transformation, the
Dirac operator is the inverse of the Euclidean propagator of
fermions). The spectral action principle then asserts that $D$ is
all that is needed to define the bosonic part of the action.
Moreover since disjoint union of spaces correspond to direct sums of
the Dirac operators, a simple additivity requirement of the action
functional shows that it has to be of the form
\begin{equation}\label{spectralaction}
S=\,\Tr(f(D/\Lambda))
\end{equation}
where $f$ is an even function of the real variable and $\Lambda$ a
parameter fixing the mass scale. In fact the choice of the test
function $f$ only plays a small role since when expanded in inverse
powers of $\Lambda$ the action $S$ only depends on the first moments
$\int\,f(u)\,u^{k-1}\,du$ and the Taylor expansion of $f$ at $0$.

\smallskip
Not surprisingly the gravitational Einstein action appears naturally
in the expansion of $S$, a point which is reminiscent of the idea of
induced gravity. Moreover in the presence of gauge fields $A$  the
operator $D$ gets modified  (replacing derivatives by covariant
derivatives) to $D_A$ and the Yang-Mills action functional ${\rm
YM}(A)$ appears, in its Euclidean form and with the correct sign if
$f\geq 0$, in the coefficient of $\Lambda^0$ in the spectral action
\eqref{spectralaction} for the operator $D_A$. The simple idea
developed in \cite{cc1}, then, is that one should understand the
modification $D\to D_A$ coming from the presence of gauge fields as
a slight change in the metric, while the action principle
\eqref{spectralaction} which is in essence purely gravitational
delivers when applied to $D_A$ the combined Einstein-Yang-Mills
action. To keep track of the ``change of metric" coming from the
gauge fields one needs to enhance the algebra of coordinates on the
manifold $M$ to the algebra of matrix valued functions on $M$ which
encodes the gauge group as its group of inner automorphisms. We
refer to section 2 of \cite{cc2} for the case of an $\SU(n)$ theory.

\smallskip
The advantage in passing to noncommutative algebras of coordinates
$\cA$ is that their automorphism group $\Aut(\cA)$ admits a
decomposition
$$
1\rightarrow\Inn(\cA)\rightarrow \Aut(\cA)\rightarrow
\Out(\cA)\rightarrow 1
$$
into inner and outer parts, which fits very well with the physics
distinction between the internal symmetries $g\in\mathcal G$  and
the others, \ie the exact sequence governing the structure of the
symmetry group $\mathcal U$ of the combined Lagrangian of gravity
and matter,
$$
1\rightarrow\mathcal G\rightarrow \mathcal U\rightarrow
\Diff(M)\rightarrow 1
$$

\smallskip
Moreover a similar decomposition into an ``inner" piece and an outer
one holds at the level of the noncommutative metric \ie of the Dirac
operator. Thus in the noncommutative world, the metrics (encoded by
$D$) admit natural ``inner fluctuations" which come directly from
the self Morita equivalence $\cA\sim\cA$ and are encoded by gauge
potentials \ie operators of the form
$$
A=\sum \,a_j\,[D,b_j]\,,\quad a_j\,,\;b_j\in \cA\,,\quad A=A^*
$$
The main result of our joint work \cite{cc1}, \cite{cc2} is that,
when applied to the inner fluctuations of the product geometry
$M\times F$  the spectral action gives the standard model coupled
with gravity. Here $M$ is a Riemannian compact spin $4$-manifold,
the  standard model coupled with gravity is in the Euclidean form,
and the geometry of the finite space $F$ is encoded (as in the
general framework of NCG) by a spectral triple $(\cA_F,\cH_F,D_F)$
\ie by a Hilbert space $\cH_F$, a representation of the algebra of
coordinates $\cA_F$, and the inverse line element $D_F$. Besides a
$\Z/2$ grading $\gamma$ this spectral triple has a crucial piece of
structure: a real structure (\cf \cite{Coreal}) \ie an antilinear
isometry of $\cH$ of square $\pm 1$ with simple algebraic rules and
whose dimension, called the $KO$-dimension is well defined modulo
$8$ from the signs involved in the algebraic rules (\cf  Appendix
\ref{appenreal}).

\smallskip
For the noncommutative geometry $F$ used in \cite{cc2} to obtain the
standard model coupled to gravity, all the ingredients are {\em
finite dimensional}. The algebra $\cA_F = \C \oplus \H \oplus M_3
(\C)$  (\ie the direct sum of the algebras $\C$ of complex numbers,
$\H$ of quaternions, and $M_3 (\C)$ of $3\times 3$ matrices) encodes
the gauge group. The Hilbert space $\cH_F$ is of dimension
$90$\footnote{It is $96$ in the model described below} and encodes
the elementary quarks and leptons. The operator $D_F$ encodes those
free parameters of the standard model related to the Yukawa
couplings.

\smallskip
For $M$ the spectral triple is given by the representation of the
algebra of smooth functions acting by multiplication in the Hilbert
space $L^2(M,S)$ of square integrable spinors, the grading is given
by $\gamma_5$ and the real structure $J_M$ is given by charge
conjugation.

\smallskip
While it is certainly remarkable to obtain the standard model action
from simple geometric principles the above work has several
shortcomings:
\begin{enumerate}
  \item The finite geometry $F$ is put in ``by hand" with no
  conceptual understanding of the representation of $\cA_F$ in
  $\cH_F $.
  \item There is a fermion doubling problem (\cf \cite{lizzi}) in
  the fermionic part of the action.
  \item It does not incorporate the neutrino mixing and see-saw
  mechanism for neutrino masses.
\end{enumerate}

\smallskip
We shall show in this note how to solve these three problems (the
first only partly since the number of generations is put by hand)
simply by keeping the distinction between the following two notions
of dimension of a noncommutative space,

\smallskip

\begin{itemize}
  \item The metric dimension
  \item The $KO$-dimension
\end{itemize}

\smallskip

The metric dimension manifests itself by the growth of the spectrum
of the Dirac operator. As far as space-time goes it appears that the
 situation of
interest will be the $4$-dimensional one. In particular the metric
dimension of the finite geometry $F$ will be zero.

\smallskip

The $KO$-dimension is only well defined modulo $8$  and it takes
into account both the $\Z/2$-grading $\gamma$ of $\cH$ as well as
the real structure $J$ (\cf Appendix \ref{appenreal}). The real
surprise is that in order for things to work the only needed change
(besides the easy addition of a right handed neutrino) is to change
the $\Z/2$ grading of the finite geometry $F$ to its opposite in the
``antiparticle" sector.
 It is only thanks to this that the Fermion doubling
problem pointed out in \cite{lizzi} can be successfully handled.
Moreover it will automatically generate the {\em full standard
model} \ie the model with neutrino mixing and the see-saw mechanism
as follows from the full classification of Dirac operators: Theorem
\ref{diracclassF}.

\smallskip

When one looks at the table \eqref{realstr} of  Appendix
\ref{appenreal} giving the $KO$-dimension of the finite space $F$
one then finds that its $KO$-dimension is now equal to $6$ modulo
$8\,$(!). As a result we see that the $KO$-dimension of the product
space $M\times F$ is in fact equal to $10\sim 2$ modulo $8$.  Of
course the above $10$ is very reminiscent of
 string theory, in which the finite space $F$ might be a good candidate
 for an ``effective" compactification at least for low energies\footnote{Note
 however that we are dealing with the
 standard model, not its supersymmetrized version.}. But $10$ is also
 $2$ modulo $8$
which might be related to the observations of \cite{reuter} about
gravity.

\smallskip It is also remarkable that the noncommutative spheres
arising from quantum groups, such as the Podle\'s spheres already
exhibit the situation where the metric dimension ($0$ in that case)
is distinct from the $KO$-dimension ($2$ in that case) as pointed
out in \cite{dab}.

\smallskip  We have gathered the definitions of the basic notions of
noncommutative geometry: spectral triples, real structure and inner
fluctuations, in the Appendix \ref{appenreal}. We shall often refer
to these basic notions in the text and urge the reader unfamiliar
with these to start by a brief look at the appendix.

\bigskip

\section{ The finite non commutative geometry
$F$}\label{finitegeom}\hfill \medskip

In this section we shall first describe in a conceptual manner
 the representation of $\cA_F$ in $\cH_F$ and classify the
Dirac operators $D_F$. The only small nuance with \cite{CoSM} is
that we incorporate a right handed neutrino $\nu_R$ and change the
$\Z/2$ grading in the antiparticle sector to its opposite. This,
innocent as it looks, allows for a better conceptual understanding
of the representation of $\cA_F$ in $\cH_F$ and also  will
completely alter the classification of Dirac operators (Theorem
\ref{diracclassF}).

\medskip

\subsection{ The representation of $\cA_F$ in
$\cH_F$}\label{therep1}\hfill \medskip

We start from the involutive algebra (with $\H$ the quaternions with
involution $q\to \bar q$)
\begin{equation}\label{algebralr}
\cA_{LR}=\,\C\oplus \H_L\oplus \H_R\oplus M_3(\C)
\end{equation}
We are looking for a natural representation
$(\cA_{LR},\cH_F,J_F,\gamma_F)$ fulfilling definition \ref{realstr}
of Appendix \ref{appenreal} in dimension $6$ modulo $8$. The
commutation relation \eqref{comm-rule} of definition \ref{realstr}
shows that there is an underlying structure of $\cA_{LR}$-bimodule
on $\cH_F$ and we shall use that structure as a guide. One uses the
bimodule structure to define ${\rm Ad}(u)$, for $u\in \cA$ unitary,
 by
 \begin{equation}\label{adjbimo}
 {\rm Ad}(u)\xi=u\xi u^*
 \end{equation}

\begin{defn} Let $\cM$ be an $\cA_{LR}$-bimodule. Then $\cM$ is
 {\em odd} iff   the adjoint action \eqref{adjbimo} of
$s=(1,-1,-1,1)$ fulfills ${\rm Ad}(s)=-1$.
\end{defn}

Such a bimodule is  a representation of the reduction of
$\cA_{LR}\otimes_\R \cA_{LR}^0$   by the projection $\frac
12\,(1-s\otimes s^0)$. This subalgebra is an algebra over $\C$ and
we restrict to complex representations. One defines the
contragredient bimodule of a bimodule $\cM$ as the complex conjugate
space
\begin{equation}\label{oppbim}
\cM^0=\{\bar \xi\;;\;\xi\in \cM\}\,,\quad
a\,\bar\xi\,b=\,\overline{b^*\xi\,a^*}\qqq \,a\,,\;b\in \cA_{LR}
\end{equation}

We can now give the following characterization of the
$\cA_{LR}$-bimodule $\cM_F$ and the real structure $J_F$ for one
generation.

\begin{prop}
\begin{itemize}
  \item The $\cA_{LR}$-bimodule $\cM_F$  is the direct sum
of all inequivalent irreducible odd $\cA_{LR}$-bimodules.
\item The dimension of $\cM_F$  is $32$.
  \item The real structure $J_F$ is given by  the isomorphism with the
contragredient bimodule.
\end{itemize}
\end{prop}

\medskip
We define the $\Z/2$-grading $\gamma_F$   by
\begin{equation}\label{gammafc}
 \gamma_F =\,c\,-\,J_F\,c\,J_F \,,\quad
c=(0,-1,1,0)\in \cA_{LR}
\end{equation}
One then checks that the following holds
\begin{equation}\label{sixmodeight}
J_F^2=1\,,\quad J_F\,\gamma_F=-\,\gamma_F\,J_F
\end{equation}
which together with the commutation of $J_F$ with the Dirac
operators, is characteristic of $KO$-dimension equal to $6$ modulo
$8$ (\cf Appendix \ref{appenreal}, definition \ref{realstr}).

The equality $\iota(\lambda,q,m)=(\lambda,q,\lambda,m)$
  defines a homomorphism $\iota$  of involutive algebras from $\cA_F$ to $\cA_{LR}$
  so that we view $\cA_F$ as a subalgebra of $\cA_{LR}$.

  \begin{defn} The real representation $(\cA_F,\cH_F,J_F,\gamma_F)$
  is the restriction to $\cA_F\subset \cA_{LR}$ of the direct sum
  $\cM_F\otimes \C^3$ of three copies of $\cM_F$.
  \end{defn}

  \medskip
  It has dimension $32\times 3=96$, needless to say this $3$ is the number of generations and it is
  put in by hand here.

\medskip

\subsection{ The unimodular unitary group $\SU(\cA_F)$}\label{therep2}\hfill \medskip

Using the action of $\cA_F$ in $\cH_F$ one defines the unimodular
subgroup $\SU(\cA_F)$ of the unitary group ${\rm U}(\cA_F)=\{u\in
\cA_F\,,\;uu^*=u^*u=1\}$ as follows,

\begin{defn} We let $\SU(\cA_F)$ be the subgroup of ${\rm U}(\cA_F)$
defined by
$$
\SU(\cA_F)=\{u\in {\rm U}(\cA_F)\;:\;{\rm Det}(u)=1\}
$$
where ${\rm Det}(u)$ is the determinant of the action of $u$ in
$\cH_F$.
\end{defn}

\medskip

One obtains both the standard model gauge group and its action on
fermions from the adjoint action of $\SU(\cA_F)$ in the following
way:

\smallskip
\begin{prop}\label{smhypercharges}
\begin{enumerate}
  \item The group $\SU(\cA_F)$ is, up to an abelian finite group, $$
  \SU(\cA_F)\sim {\rm U}(1)\times
\SU(2)\times\SU(3)$$
  \item The adjoint action $u\to {\rm Ad}(u)$  (\cf \eqref{adjbimo})  of $\SU(\cA_F)$ in $\cH_F$ coincides with the
standard model action on elementary quarks and leptons.
\end{enumerate}
 \end{prop}

To see what happens we first have to label the basis of the bimodule
$\cM_F$. We use the following idempotents in $\cA_{LR}$,
\begin{equation}\label{idempotents}
e_\ell=(1,0,0,0)\,,\quad e_L=(0,1,0,0)\,,\quad e_R=(0,0,1,0)\,,\quad
e_q=(0,0,0,1)\,.
\end{equation}
The reduced algebras are respectively $\C$, $\H_L$, $\H_R$,
$M_3(\C)$. One has $\sum\,e_j=1$ and similarly in the algebra
$\cA_{LR}\otimes \cA_{LR}^0$ one has $$ \sum \,e_j\otimes e_k^0=\,1
$$
 Using the action of $\cA_{LR}\otimes
\cA_{LR}^0$ associated to the bimodule structure of $\cM$ thus gives
a decomposition of the form
$$
\cM=\,\sum\,e_j\,\cM\,e_k
$$
Since $\cM$ is odd this decomposition can be written as
$$
\cM=\,\sum\,e_j\,\cM\,e_K +\,\sum\,e_J\,\cM\,e_k
$$
Let us consider the term $e_L\,\cM\,e_\ell $. It is a $\H_L$-left,
$\C$-right module. Thus it is a multiple of the only irreducible
representation $\pi_L$ of $\H_L$ which is two dimensional. The
action of $\C$ is given by the scalar action. Let us consider the
term $e_L\,\cM\,e_q $. It is a $\H_L$-left, $M_3(\C)$-right module.
Since the algebra $\H_L\otimes_\R M_3(\C)$ is $M_6(\C)$ we see that
all such bimodules are multiples of the bimodule $\pi_L^3$ given for
the left action of $\H_L$ as the direct sum of three copies of
$\pi_L$ and with the obvious right action of $M_3(\C)$ permuting the
three copies. Exactly the same holds for the bimodules
$e_R\,\cM\,e_\ell $ and $e_R\,\cM\,e_q $, which are respectively
multiples of $\pi_R$ and of $\pi_R^3$.
 Similar results hold switching the left and right actions
 \ie by passing to the contragredient bimodule of $\cM$.
We thus see that the sum of the irreducible odd bimodules is given
by
\begin{equation}\label{irredbim}
\cM_F=(\pi_L\oplus\pi_R\oplus \pi_R^3\oplus
\pi_L^3)\,\oplus\,(\pi_L\oplus\pi_R\oplus \pi_R^3\oplus \pi_L^3)^0
\end{equation}
This $\cA_{LR}$-bimodule $\cM_F$ is of dimension $2\cdot(2+2+2\times
3+2\times 3)=32$ and the adjoint action gives the gauge action of
the standard model  for one generation, with the following labels
for the basis elements of $\cM_F$,
$$
\left(
       \begin{array}{cc}
         \nu_L  & \nu_R  \\
         e_L & e_R \\
       \end{array}
     \right)
     $$
     for the term $\pi_L\oplus\pi_R$,
     $$
\left(\begin{array}{cc}
   u_L^j & u_R^j \\
    d_L^j & d_R^j \\
    \end{array}
      \right)
    $$
    for the term $\pi_R^3\oplus
\pi_L^3$ ( with color indices $j$) and the transformation $q\to \bar
q$ to pass to the contragredient bimodules. With these labels one
checks that
 the  adjoint action  of the ${\rm U}(1)$
  factor is given by multiplication of the basis vectors $f$
  by the following powers of $\lambda\in {\rm U}(1)$:
$$
\begin{matrix}\hbox{~~~~~}&e &\nu &u &
d \\
&&&&\cr
f_L&-1&-1&\frac{1}{3}&\frac{1}{3}\\
&&&&\\
f_R&-2&0&\frac{4}{3}&-\frac{2}{3}\end{matrix}
$$

\medskip

\subsection{ The classification of Dirac operators}\label{therep3}\hfill \medskip

To be precise we adopt the following,

\begin{defn} A Dirac operator is a self-adjoint operator $D$
in $\cH_F$  commuting with $J_F$, $\C_F=\{(\lambda,\lambda,0)\}\in
\cA_F $, anticommuting with $\gamma_F$ and fulfilling the order one
condition $[[D,a],b^0] = 0$ for any $a,b \in \cA_F$.
\end{defn}

\smallskip
The physics meaning of the condition of commutation with $\C_F$ is
to ensure that one gauge vector boson (the photon) remains massless.

\medskip
In order to state the classification of Dirac operators we introduce
the following notation, let $M_e$, $M_\nu$, $M_d$, $M_u$ and $M_R$
be three by three matrices, we then let $D(M)$ be the operator in
$\cH_F$ given by
\begin{equation}\label{dofm}
D(M)=\left[\begin{matrix}S &T^*\\
T &\bar S\end{matrix} \right]
\end{equation}
where
\begin{equation}\label{sop}
S = S_{\ell} \, \oplus (S_q \otimes 1_3)
\end{equation}
and in the basis $(\nu_R,e_R,\nu_L,e_L)$ and $(u_R,d_R,u_L,d_L)$,
\medskip
\begin{equation}\label{sops}
S_\ell = \left[\begin{matrix}0&0&M^{*}_\nu&0\\
0&0&0&M^{*}_e\\
M_\nu&0&0&0\\
0&M_e&0&0&\end{matrix} \right]\qquad
S_q = \left[\begin{matrix}0&0&M^{*}_u&0\\
0&0&0&M^{*}_d\\
M_u&0&0&0\\
0&M_d&0&0&\end{matrix} \right]
\end{equation}

\medskip
while the operator $T$ is $0$ except on the
 subspace $\cH_{\nu_R}\subset \cH_F$ with basis the $\nu_R$ which
 it maps,  using the matrix $M_R$, to the subspace $\cH_{\bar\nu_R}\subset \cH_F$
 with basis the $\bar\nu_R$.

\medskip
\begin{thm}\label{diracclassF}
\begin{enumerate}
\item Let $D$ be a Dirac operator. There exists three by three matrices $M_e$, $M_\nu$, $M_d$, $M_u$
and $M_R$, with $M_R$   symmetric, such that $D=D(M)$.
\item  All operators
$D(M)$ (with $M_R$   symmetric) are Dirac operators.
     \item The operators $D(M)$ and $D(M')$ are conjugate by a unitary  operator
  commuting with $\cA_F$, $\gamma_F$ and $J_F$ iff there exists
  unitary matrices $V_j$ and $W_j$ such that
  $$
  M'_e=V_1\,M_e\,V_3^*\,,\; M'_\nu=V_2\,M_\nu\,V_3^*\,,\;
  M'_d=W_1\,M_d\,W_3^*\,,\; M'_u=W_2\,M_u\,W_3^*\,,\; M'_R=V_2\,M_R\,\bar V_2^*
  $$
  \end{enumerate}
\end{thm}

In particular Theorem \ref{diracclassF} shows that
 the Dirac operators  give all the required features, such as
 \begin{itemize}
   \item Mixing matrices for quarks and leptons
   \item Unbroken color
   \item See-saw mechanism for right handed neutrinos
 \end{itemize}

\medskip
Let us briefly explain the last item, \ie the analogue of the seesaw
mechanism in our context. The restriction of $D(M)$ to the subspace
of $\cH_F$ with basis the $(\nu_R,\nu_L,\bar\nu_R,\bar\nu_L)$ is
given by the matrix,
\begin{equation}\label{seesawmatrix}
\left[
  \begin{array}{cccc}
    0 & M_\nu^* & M_R^* & 0 \\
    M_\nu & 0 & 0 & 0 \\
    M_R & 0 & 0 & \bar{M}_\nu^* \\
    0 & 0 & \bar{M}_\nu & 0 \\
  \end{array}
\right]
\end{equation}
Let us simplify to one generation and let $M_R\sim M$ be a very
large mass term- the largest eigenvalue of $M_R$ will be set to the
order of the unification scale by the equations of motion
\eqref{mrstarmr} of the spectral action below- while $M_\nu\sim m$
is much smaller\footnote{it is a Dirac mass term, fixed by the Higgs
vev}. The eigenvalues of the matrix \eqref{seesawmatrix} are then
given by
$$
\frac 12\,(\pm M\, \pm \sqrt{M^2+4m^2})
$$
which gives two eigenvalues very close to  $\pm M$ and two others
very close to $\pm \frac{m^2}{M}$ as can be checked directly from
the determinant of the matrix \eqref{seesawmatrix}, which is equal
to $|M_\nu|^4\sim m^4$ (for one generation).

\medskip
\section{The spectral action for $M\times F$ and the Standard Model}\label{NCGSMinputSect}\hfill\medskip

We now consider a 4-dimensional smooth compact Riemannian manifold
$M$ with a fixed spin structure and recall that it is fully encoded
by its Dirac spectral triple
$(\cA_1,\cH_1,D_1)=(C^\infty(M),L^2(M,S),\dirac_M)$. We then
consider its product with the above finite geometry
$(\cA_2,\cH_2,D_2)=(\cA_F,\cH_F,D_F)$. With $(\cA_j,\cH_j,D_j)$ of
$KO$-dimensions $4$ for $j=1$ and $6$ for $j=2$, the product
geometry is given by the rules,
$$
\cA = \cA_1 \otimes \cA_2 \ , \ \cH = \cH_1 \otimes \cH_2 \ , \ D =
D_1 \otimes 1 + \gamma_1 \otimes D_2 \, , \ \gamma=\gamma_1\otimes
\gamma_2\,,  \ J=J_1\otimes J_2
$$
Note that it matters that $J_1$ commutes with $\gamma_1$ to check
that $J$ commutes with $D$.
 The $KO$-dimension of the
finite space $F$ is $6\in \Z/8$ and thus the $KO$-dimension of the
product geometry $M\times F$  is now $2\in \Z/8$. In other words
according to Appendix \ref{appenreal}, definition \ref{realstr} the
commutation rules are
\begin{equation}\label{per82}
J^2 = -1, \ \ \ \ JD =   DJ, \ \ \text{and} \ \ J\gamma = - \gamma J
\, .
\end{equation}
Let us now explain how these rules allow to define a natural
antisymmetric bilinear form on the even part $\cH^+=\{\xi\in
\cH\;,\;\gamma\,\xi=\xi\}$ of $\cH$.

\smallskip
\begin{prop} On a real spectral triple of $KO$-dimension $2\in
\Z/8$, the following equality defines an antisymmetric bilinear form
on $\cH^+=\{\xi\in \cH\;,\;\gamma\,\xi=\xi\}$,
\begin{equation}\label{antisybil}
A_D(\xi',\xi)=\,\langle\,J\,\xi',D\,\xi\rangle \qqq \xi, \xi'\in
\cH^+
\end{equation}
The above trilinear pairing between $D$, $\xi$ and $\xi'$ is gauge
invariant under the adjoint action (\cf \eqref{adjact}) of the
unitary group of $\cA$,
\begin{equation}\label{antisybil1}
A_D(\xi',\xi)=A_{D_u}({\rm Ad}(u)\xi',{\rm Ad}(u)\xi)\,,\quad
D_u={\rm Ad}(u)\,D\,{\rm Ad}(u^*)
\end{equation}
\end{prop}

\medskip

Now the Pfaffian of an antisymmetric bilinear form is best expressed
in terms of the functional integral involving anticommuting
``classical fermions" which at the formal level means that
$$
{\rm Pf}(A)=\int\,e^{-\frac 12\,A(\xi)}\,D[\xi]
$$
It is the use of the Pfaffian as a square root of the determinant
that allows to solve the Fermion doubling puzzle which was pointed
out in \cite{lizzi}. The solution obtained by a better choice of the
$KO$-dimension of the space $F$ and hence of $M\times F$ is not
unrelated to  the point made in \cite{gbis}.

\medskip

\begin{thm} Let $M$ be a Riemannian spin $4$-manifold and $F$ the finite noncommutative geometry of $KO$-dimension $6$ described above.
Let $M\times F$ be endowed with the product metric.
\begin{enumerate}
  \item The unimodular subgroup of the unitary group acting by the adjoint
  representation ${\rm Ad}(u)$ in $\cH$ is the group of
 gauge transformations of SM.
  \item The unimodular inner fluctuations $A$ of the metric (\cf Appendix \ref{appenreal}) are parameterized exactly by the
  gauge bosons of SM (including the Higgs doublet).
  \item The full standard
  model (see the explicit formula in \S \ref{fullsm}) minimally coupled with Einstein gravity
  is given in Euclidean form by the action functional\footnote{We
  take $f$ even and positive with $f^{(n)}(0)=0$ for $n\geq 1$ for
  definiteness. Note also that the components of $\xi$ anticommute so the
  antisymmetric form does not vanish.}
  $$
S=\,\Tr(f(D_A/\Lambda))+\frac
12\,\langle\,J\,\xi,D_A\,\xi\rangle\,,\quad\xi\in \cH^+
  $$
  applied to unimodular  inner fluctuations $D_A=D+A+JAJ^{-1}$ of the metric.
\end{enumerate}
\end{thm}

\medskip
The proof is an excruciating computation, which is a variant of
\cite{cc2} (\cf \cite{kastler} for a detailed version). After
turning off gravity to simplify and working in flat space (after
Wick rotation back to Lorentzian signature) one gets exactly the
Lagrangian
 of \S \ref{fullsm} which can hardly be fortuitous.
The fermion doubling problem is resolved by the use of the Pfaffian,
we checked that part for the Dirac mass terms, and trust that the
same holds for the Majorana mass terms. There is one subtle point
which is the use of the following chiral transformation:
$$
U=\,e^{i\frac{\pi}{4}\gamma_5}
$$
to transform the fermionic part of the action  to the traditional
one \ie the Euclidean action for Fermi fields (\cf \cite{Coleman}).
While this transformation is innocent at the classical level, it is
non-trivial at the quantum level and introduces some kind of Maslov
index in the transition from our form of the Euclidean action to the
more traditional one. We shall now give more details on the bosonic
part of the action.

\smallskip

\section{Detailed form of the Bosonic action}\label{bosact}

We shall now give the precise form of the bosonic action, the
calculation is entirely similar to \cite{cc2} with new terms
appearing from the presence of $M_R$.

\smallskip

One lets $f_k=\int_0^\infty\,f(u)\,u^{k-1}du$ for $k>0$ and
$f_0=f(0)$. Also
\begin{eqnarray}
  a &=& \,\Tr(M_\nu^*M_\nu+M_e^*M_e+3(M_u^*M_u+M_d^*M_d)) \label{momentslabels}\\
  b &=& \,\Tr((M_\nu^*M_\nu)^2+(M_e^*M_e)^2+3(M_u^*M_u)^2+3(M_d^*M_d)^2) \nonumber \\
  c &=& \Tr(M_R^*M_R)\nonumber  \\
  d &=& \Tr((M_R^*M_R)^2) \nonumber \\
  e &=& \Tr(M_R^*M_R M_\nu^*M_\nu)\nonumber
\end{eqnarray}

 The spectral  action is given by a computation entirely similar to
 \cite{cc2} which yields:
\begin{eqnarray}
  S &=& \, \frac{1}{\pi^2}(48\,f_4\,\Lambda^4-f_2\,\Lambda^2\,c+\frac{
f_0}{4}\,d)\,\int \,\sqrt g\,d^4 x \label{bossm}\\
    &+& \, \frac{96\,f_2\,\Lambda^2 -f_0\,c}{ 24\pi^2} \, \int\,R
 \, \sqrt g \,d^4 x  \nonumber\\
    &+& \, \frac{f_0 }{ 10\,\pi^2} \int\,(\frac{11}{6}\,R^* R^* -3 \, C_{\mu
\nu \rho \sigma} \, C^{\mu \nu \rho \sigma})\, \sqrt g \,d^4 x \nonumber \\
 &+&  \, \frac{(- 2\,a\,f_2
  \,\Lambda^2\,+ \,e\,f_0 )}{ \pi^2} \int\,  |\varphi|^2\, \sqrt g \,d^4 x \nonumber \\
    &+&  \, \frac{f_0 }{ 2\,\pi^2} \int\, a\, |D_{\mu} \varphi|^2\, \sqrt g \,d^4 x \nonumber \\
   &-&  \frac{f_0}{ 12\,\pi^2} \int\, a \,R \, |\varphi|^2 \, \sqrt g \,d^4 x
 \nonumber\\
    &+& \, \frac{f_0 }{ 2\,\pi^2} \int\,(g_{3}^2 \, G_{\mu \nu}^i \, G^{\mu \nu i} +  g_{2}^2 \, F_{\mu
\nu}^{ \alpha} \, F^{\mu \nu  \alpha}+\, \frac{5}{ 3} \,
g_{1}^2 \,  B_{\mu \nu} \, B^{\mu \nu})\, \sqrt g \,d^4 x\nonumber \\
    &+&\, \frac{f_0 }{ 2\,\pi^2} \int\,b\, |\varphi|^4 \, \sqrt g \,d^4 x \nonumber
\end{eqnarray}
where $(a,b,c,d,e)$ are defined above and $D_\mu\varphi$ is the
minimal coupling. A simple change of variables as in  \cite{cc2},
namely
\begin{equation}\label{rescalehiggs}
\higgs   =\frac{\sqrt{a\,f_0}}{ \pi}  \, \varphi \, ,
\end{equation}
so that the kinetic term becomes\footnote{here we differ slightly
from \cite{cc2} by a factor of $\sqrt 2$ to match the conventions of
Veltman \cite{VDiag}}
$$
\int\,\frac 12|D_\mu \higgs|^2\,\sqrt g \,d^4x
$$
and
\begin{equation}\label{coeffymterm}
  \frac{g_{3}^2 \, f_0 }{ 2\pi^2} =\frac 14\,,\quad
  g_{3}^2=\ g_{2}^2 = \frac{5}{ 3} \, g_{1}^2 \, .
\end{equation}
 transforms the bosonic action
 into the form:
\begin{eqnarray}
  S &=& \ \int d^4 x \, \sqrt g \ \biggl[ \frac{1}{
2\kappa_0^2} \, R + \alpha_0 \, C_{\mu \nu \rho \sigma}
\, C^{\mu \nu \rho \sigma} \label{rescaledbosaction}\\
    &+& \ \ \gamma_0 + \tau_0 \, {}^* R {}^*
R + \delta_0 \, R;_{\mu} {}^{\mu} \nonumber\\
    &+&  \frac{1}{ 4} \, G_{\mu
\nu}^i \, G^{\mu \nu i} + \frac{1}{ 4} \, F_{\mu \nu}^{ \alpha} \,
F^{\mu \nu  \alpha}+\ \frac{1}{ 4} \, B_{\mu \nu}\, B^{\mu
\nu} \nonumber\\
    &+&   \,\frac 12\vert D_{\mu} \, \higgs\vert^2 - \mu_0^2 |\higgs|^2 - \frac{1}{ 12}
\, R \,\vert \higgs \vert^2 + \lambda_0 |\higgs|^4 \biggl]\nonumber
\end{eqnarray}

where
\begin{eqnarray}
\frac{1}{\kappa_0^2} &=& \ \frac{96\,f_2\,\Lambda^2
-f_0\,c}{ 12\,\pi^2} \\
 \mu_0^2 &=& \ 2\,\frac{f_2\,\Lambda^2}{ f_0}-\,\frac{e}{a} \\
  \alpha_0 &=& -\frac{3\,f_0 }{ 10\,\pi^2} \\
   \tau_0  &=& \frac{11\,f_0 }{ 60\,\pi^2} \\
  \delta_0  &=& \ -\frac{2}{ 3} \, \alpha_0  \\
  \gamma_0 &=& \ \frac{1}{\pi^2}(48\,f_4\,\Lambda^4-f_2\,\Lambda^2\,c+\frac{
f_0}{4}\,d) \\
  \lambda_0  &=& \frac{\pi^2 }{2\,
f_0}\frac{b}{ a^2}=\frac{b\,g^2}{a^2}
 \end{eqnarray}

\bigskip

\section{Detailed form of the spectral action without gravity}\label{fullsm}

To make the comparison easier we Wick rotate back to Minkowski space
and after turning off gravity by working in flat space (and addition
of gauge fixing terms\footnote{We add the Feynman gauge fixing terms
just to simplify the form of the gauge kinetic terms}) the spectral
action,  after the change of variables summarized in table
\ref{smtospec},  is given by the following formula:

\bigskip
\begin{center}
\begin{math}
{\mathcal
L}_{SM}=-\frac{1}{2}\partial_{\nu}g^{a}_{\mu}\partial_{\nu}g^{a}_{\mu}
-g_{s}f^{abc}\partial_{\mu}g^{a}_{\nu}g^{b}_{\mu}g^{c}_{\nu}
-\frac{1}{4}g^{2}_{s}f^{abc}f^{ade}g^{b}_{\mu}g^{c}_{\nu}g^{d}_{\mu}g^{e}_{\nu}
-\partial_{\nu}W^{+}_{\mu}\partial_{\nu}W^{-}_{\mu}-M^{2}W^{+}_{\mu}W^{-}_{\mu}
-\frac{1}{2}\partial_{\nu}Z^{0}_{\mu}\partial_{\nu}Z^{0}_{\mu}-\frac{1}{2c^{2}_{w}}
M^{2}Z^{0}_{\mu}Z^{0}_{\mu}
-\frac{1}{2}\partial_{\mu}A_{\nu}\partial_{\mu}A_{\nu}
-igc_{w}(\partial_{\nu}Z^{0}_{\mu}(W^{+}_{\mu}W^{-}_{\nu}-W^{+}_{\nu}W^{-}_{\mu})
-Z^{0}_{\nu}(W^{+}_{\mu}\partial_{\nu}W^{-}_{\mu}-W^{-}_{\mu}\partial_{\nu}W^{+}_{\mu})
+Z^{0}_{\mu}(W^{+}_{\nu}\partial_{\nu}W^{-}_{\mu}-W^{-}_{\nu}\partial_{\nu}W^{+}_{\mu}))
-igs_{w}(\partial_{\nu}A_{\mu}(W^{+}_{\mu}W^{-}_{\nu}-W^{+}_{\nu}W^{-}_{\mu})
-A_{\nu}(W^{+}_{\mu}\partial_{\nu}W^{-}_{\mu}-W^{-}_{\mu}\partial_{\nu}W^{+}_{\mu})
+A_{\mu}(W^{+}_{\nu}\partial_{\nu}W^{-}_{\mu}-W^{-}_{\nu}\partial_{\nu}W^{+}_{\mu}))
-\frac{1}{2}g^{2}W^{+}_{\mu}W^{-}_{\mu}W^{+}_{\nu}W^{-}_{\nu}+\frac{1}{2}g^{2}
W^{+}_{\mu}W^{-}_{\nu}W^{+}_{\mu}W^{-}_{\nu}
+g^2c^{2}_{w}(Z^{0}_{\mu}W^{+}_{\mu}Z^{0}_{\nu}W^{-}_{\nu}-Z^{0}_{\mu}Z^{0}_{\mu}W^{+}_{\nu}
W^{-}_{\nu})
+g^2s^{2}_{w}(A_{\mu}W^{+}_{\mu}A_{\nu}W^{-}_{\nu}-A_{\mu}A_{\mu}W^{+}_{\nu}
W^{-}_{\nu})
+g^{2}s_{w}c_{w}(A_{\mu}Z^{0}_{\nu}(W^{+}_{\mu}W^{-}_{\nu}-W^{+}_{\nu}W^{-}_{\mu})
-2A_{\mu}Z^{0}_{\mu}W^{+}_{\nu}W^{-}_{\nu})
-\frac{1}{2}\partial_{\mu}H\partial_{\mu}H-2M^2\alpha_{h}H^{2}
-\partial_{\mu}\phi^{+}\partial_{\mu}\phi^{-}
-\frac{1}{2}\partial_{\mu}\phi^{0}\partial_{\mu}\phi^{0}
-\beta_{h}\left(\frac{2M^{2}}{g^{2}}+\frac{2M}{g}H+\frac{1}{2}(H^{2}+\phi^{0}\phi^{0}+2\phi^{+}\phi^{-
})\right)
  +\frac{2M^{4}}{g^{2}}\alpha_{h}
  -g\alpha_h
M\left(H^3+H\phi^{0}\phi^{0}+2H\phi^{+}\phi^{-}\right)
-\frac{1}{8}g^{2}\alpha_{h}
\left(H^4+(\phi^{0})^{4}+4(\phi^{+}\phi^{-})^{2}
+4(\phi^{0})^{2}\phi^{+}\phi^{-}
+4H^{2}\phi^{+}\phi^{-}+2(\phi^{0})^{2}H^{2}\right)
-gMW^{+}_{\mu}W^{-}_{\mu}H-\frac{1}{2}g\frac{M}{c^{2}_{w}}Z^{0}_{\mu}Z^{0}_{\mu}H
-\frac{1}{2}ig\left(W^{+}_{\mu}(\phi^{0}\partial_{\mu}\phi^{-}
-\phi^{-}\partial_{\mu}\phi^{0})
-W^{-}_{\mu}(\phi^{0}\partial_{\mu}\phi^{+}
-\phi^{+}\partial_{\mu}\phi^{0})\right)
+\frac{1}{2}g\left(W^{+}_{\mu}(H\partial_{\mu}\phi^{-}
-\phi^{-}\partial_{\mu}H)
 +W^{-}_{\mu}(H\partial_{\mu}\phi^{+}-\phi^{+}\partial_{\mu}H)\right)
+\frac{1}{2}g\frac{1}{c_{w}}(Z^{0}_{\mu}(H\partial_{\mu}\phi^{0}-\phi^{0}\partial_{\mu}H)
+M\,(\frac{1}{c_{w}}Z^{0}_{\mu}\partial_{\mu}\phi^{0}+W^{+}_{\mu}
\partial_{\mu}\phi^{-}+W^{-}_{\mu}
\partial_{\mu}\phi^{+})
-ig\frac{s^{2}_{w}}{c_{w}}MZ^{0}_{\mu}(W^{+}_{\mu}\phi^{-}-W^{-}_{\mu}\phi^{+})
   +igs_{w}MA_{\mu}(W^{+}_{\mu}\phi^{-}-W^{-}_{\mu}\phi^{+})
-ig\frac{1-2c^{2}_{w}}{2c_{w}}Z^{0}_{\mu}(\phi^{+}\partial_{\mu}\phi^{-}
-\phi^{-}\partial_{\mu}\phi^{+})
+igs_{w}A_{\mu}(\phi^{+}\partial_{\mu}\phi^{-}-\phi^{-}\partial_{\mu}\phi^{+})
-\frac{1}{4}g^{2}W^{+}_{\mu}W^{-}_{\mu}
\left(H^{2}+(\phi^{0})^{2}+2\phi^{+}\phi^{-}\right) -\frac{1}{8}
g^{2}\frac{1}{c^{2}_{w}}Z^{0}_{\mu}Z^{0}_{\mu}
\left(H^{2}+(\phi^{0})^{2}+2(2s^{2}_{w}-1)^{2}\phi^{+}\phi^{-}\right)
-\frac{1}{2}g^{2}\frac{s^{2}_{w}}{c_{w}}Z^{0}_{\mu}\phi^{0}(W^{+}_{\mu}\phi^{-}+W^{-}_{\mu}\phi^{+})
-\frac{1}{2}ig^{2}\frac{s^{2}_{w}}{c_{w}}Z^{0}_{\mu}H(W^{+}_{\mu}\phi^{-}-W^{-}_{\mu}\phi^{+})
+\frac{1}{2}g^{2}s_{w}A_{\mu}\phi^{0}(W^{+}_{\mu}\phi^{-}+W^{-}_{\mu}\phi^{+})
+\frac{1}{2}ig^{2}s_{w}A_{\mu}H(W^{+}_{\mu}\phi^{-}-W^{-}_{\mu}\phi^{+})
-g^{2}\frac{s_{w}}{c_{w}}(2c^{2}_{w}-1)Z^{0}_{\mu}A_{\mu}\phi^{+}\phi^{-}
-g^{2}s^{2}_{w}A_{\mu}A_{\mu}\phi^{+}\phi^{-}
+\frac 12 i
g_s\,\lambda_{ij}^a(\bar{q}^{\sigma}_{i}\gamma^{\mu}q^{\sigma}_{j})g^{a}_{\mu}
-\bar{e}^{\lambda}(\gamma\partial+m^{\lambda}_{e})e^{\lambda}
-\bar{\nu}^{\lambda}(\gamma\partial+m^{\lambda}_{\nu})\nu^{\lambda}
-\bar{u}^{\lambda}_{j}(\gamma\partial+m^{\lambda}_{u})u^{\lambda}_{j}
-\bar{d}^{\lambda}_{j}(\gamma\partial+m^{\lambda}_{d})d^{\lambda}_{j}
+igs_{w}A_{\mu}\left(-(\bar{e}^{\lambda}\gamma^{\mu}
e^{\lambda})+\frac{2}{3}(\bar{u}^{\lambda}_{j}\gamma^{\mu}
u^{\lambda}_{j})-\frac{1}{3}(\bar{d}^{\lambda}_{j}\gamma^{\mu}
d^{\lambda}_{j})\right)
   +\frac{ig}{4c_{w}}Z^{0}_{\mu}
\{(\bar{\nu}^{\lambda}\gamma^{\mu}(1+\gamma^{5})\nu^{\lambda})+
(\bar{e}^{\lambda}\gamma^{\mu}(4s^{2}_{w}-1-\gamma^{5})e^{\lambda})
     +(\bar{d}^{\lambda}_{j}\gamma^{\mu}(\frac{4}{3}s^{2}_{w}-1-\gamma^{5})d^{\lambda}_{j})+
(\bar{u}^{\lambda}_{j}\gamma^{\mu}(1-\frac{8}{3}s^{2}_{w}+\gamma^{5})u^{\lambda}_{j})
\}
+\frac{ig}{2\sqrt{2}}W^{+}_{\mu}\left((\bar{\nu}^{\lambda}\gamma^{\mu}(1+\gamma^{5})\cx_{\lambda\kappa}e^{\kappa})
+(\bar{u}^{\lambda}_{j}\gamma^{\mu}(1+\gamma^{5})C_{\lambda\kappa}d^{\kappa}_{j})\right)
+\frac{ig}{2\sqrt{2}}W^{-}_{\mu}\left((\bar{e}^{\kappa}\cx^{\dagger}_{\kappa\lambda}\gamma^{\mu}(1+\gamma^{5})\nu^{\lambda})
+(\bar{d}^{\kappa}_{j}C^{\dagger}_{\kappa\lambda}\gamma^{\mu}(1+\gamma^{5})u^{\lambda}_{j})\right)
   +\frac{ig}{2M\sqrt{2}}\phi^{+}
\left(-m^{\kappa}_{e}(\bar{\nu}^{\lambda}\cx_{\lambda\kappa}(1-\gamma^{5})e^{\kappa})
+m^{\lambda}_{\nu}(\bar{\nu}^{\lambda}\cx_{\lambda\kappa}(1+\gamma^{5})e^{\kappa}\right)
   +\frac{ig}{2M\sqrt{2}}\phi^{-}
\left(m^{\lambda}_{e}(\bar{e}^{\lambda}\cx^{\dagger}_{\lambda\kappa}(1+\gamma^{5})\nu^{\kappa})
-m^{\kappa}_{\nu}(\bar{e}^{\lambda}\cx^{\dagger}_{\lambda\kappa}(1-\gamma^{5})\nu^{\kappa}\right)
-\frac{g}{2}\frac{m^{\lambda}_{\nu}}{M}H(\bar{\nu}^{\lambda}\nu^{\lambda})
-\frac{g}{2}\frac{m^{\lambda}_{e}}{M}H(\bar{e}^{\lambda}e^{\lambda})
+\frac{ig}{2}\frac{m^{\lambda}_{\nu}}{M}\phi^{0}(\bar{\nu}^{\lambda}\gamma^{5}\nu^{\lambda})
-\frac{ig}{2}\frac{m^{\lambda}_{e}}{M}\phi^{0}(\bar{e}^{\lambda}\gamma^{5}e^{\lambda})
-\frac 14\,\bar
\nu_\lambda\,M^R_{\lambda\kappa}\,(1-\gamma_5)\hat\nu_\kappa -\frac
14\,\overline{\bar
\nu_\lambda\,M^R_{\lambda\kappa}\,(1-\gamma_5)\hat\nu_\kappa}
+\frac{ig}{2M\sqrt{2}}\phi^{+}
\left(-m^{\kappa}_{d}(\bar{u}^{\lambda}_{j}C_{\lambda\kappa}(1-\gamma^{5})d^{\kappa}_{j})
+m^{\lambda}_{u}(\bar{u}^{\lambda}_{j}C_{\lambda\kappa}(1+\gamma^{5})d^{\kappa}_{j}\right)
   +\frac{ig}{2M\sqrt{2}}\phi^{-}
\left(m^{\lambda}_{d}(\bar{d}^{\lambda}_{j}C^{\dagger}_{\lambda\kappa}(1+\gamma^{5})u^{\kappa}_{j})
-m^{\kappa}_{u}(\bar{d}^{\lambda}_{j}C^{\dagger}_{\lambda\kappa}(1-\gamma^{5})u^{\kappa}_{j}\right)
-\frac{g}{2}\frac{m^{\lambda}_{u}}{M}H(\bar{u}^{\lambda}_{j}u^{\lambda}_{j})
-\frac{g}{2}\frac{m^{\lambda}_{d}}{M}H(\bar{d}^{\lambda}_{j}d^{\lambda}_{j})
+\frac{ig}{2}\frac{m^{\lambda}_{u}}{M}\phi^{0}(\bar{u}^{\lambda}_{j}\gamma^{5}u^{\lambda}_{j})
-\frac{ig}{2}\frac{m^{\lambda}_{d}}{M}\phi^{0}(\bar{d}^{\lambda}_{j}\gamma^{5}d^{\lambda}_{j})
\end{math}
\end{center}

\medskip
This formula compares nicely
 with \cite{VDiag}.
 Besides the addition of the neutrino mass terms, and absence of the
ghost terms  there is only  one difference: in the spectral action
Lagrangian one gets the term:
\begin{equation}\label{golds}
M\,(\frac{1}{c_{w}}Z^{0}_{\mu}\partial_{\mu}\phi^{0}+W^{+}_{\mu}
\partial_{\mu}\phi^{-}+W^{-}_{\mu}
\partial_{\mu}\phi^{+})
\end{equation}
while in the Veltman's formula \cite{VDiag} one gets instead the
following:
\begin{equation}\label{ghos}
-M^{2}\phi^{+}\phi^{-}-\frac{1}{2c^{2}_{w}}M^2\phi^{0}\phi^{0}
\end{equation}
This difference comes from the gauge fixing term
\begin{equation}\label{feynthooft}
\cL_{fix}=-\frac{1}{2} \,\cC^2\,,\quad \cC_a=\,-\partial_\mu
W_a^\mu+M_a\,\phi_a
\end{equation}
given by the Feynman-t'Hooft gauge  in Veltman's formula
\cite{VDiag}, indeed one has
\begin{equation}\label{feynthooft1}
\cL_{fix}=-\frac{1}{2} \,(\partial_\mu
W_a^\mu)^2-\frac{1}{2c^{2}_{w}}M^2\phi^{0}\phi^{0}-M^{2}\phi^{+}\phi^{-}
+M\,(\frac{1}{c_{w}}\phi^{0}\partial_{\mu}Z^{0}_{\mu}+
\phi^{-}\partial_{\mu} W^{+}_{\mu}+\phi^{+}\partial_{\mu}W^{-}_{\mu}
)
\end{equation}

\smallskip
The numerical values are similar to those of \cite{cc2} and in
particular one gets the same value of gauge couplings as in grand
unified  $\SU(5)$-theory. This means that in the above formula the
values of $g$, $g_s$ and $s_w$, $c_w$ are fixed exactly as in
\cite{cc2} at
\begin{equation}\label{gut}
g_s=g\,,\quad {\rm tg}(w)^2=\frac 35
\end{equation}
One also gets a specific value of the Higgs scattering parameter
$\alpha_h$, as in \cite{cc2} (which agrees with \cite{knecht}),
\begin{equation}\label{higgsscat}
\alpha_h=\frac{8\,b}{a^2}
\end{equation}
(with the notations \eqref{momentslabels}) which is of the order of
$\frac 83$ if there is a dominating top mass. The change of
notations for the Higgs fields is
\begin{equation}\label{higgstranslate}
\higgs=\frac{1}{\sqrt 2}\frac{\sqrt{a}}
 {g} (1+\psi)=(\frac{2M}{g}+
H-i\phi^{0},-i\sqrt{2}\phi^{+})\,,
\end{equation}
while the huge term in $\Lambda^2$ in the spectral action can be
absorbed by a suitable choice of the tadpole constant $\beta_h$
\begin{equation}\label{tadpole}
\beta_h=2\alpha_{h}\,M^2-4\frac{f_2\Lambda^2}{f_0} +2\frac ea
\end{equation}

\smallskip
Note that the matrices $M_u$, $M_d$, $M_\nu$ and $M_e$ are only
relevant up to an overall scale. Indeed they only enter in the
coupling of the Higgs with fermions and because of the rescaling
\eqref{rescalehiggs} only by the terms
\begin{equation}\label{rescaledyukawamasses}
k_x=\,\frac{\pi}{\sqrt{a\,f_0}}\,M_x \,,\quad x\in\{u,d,\nu,e\}
\end{equation}
which are dimensionless matrices by construction. The conversion for
the mass matrices is
\begin{eqnarray}
  (k_u)_{\lambda\kappa} &=& \frac{g}{2M}\,m_u^\lambda\,\delta_\lambda^\kappa \label{massrelations}\\
  (k_d)_{\lambda\kappa} &=& \frac{g}{2M}\,m_d^\mu\,C_{\lambda\mu}\delta_\mu^\rho C^\dagger_{\rho\kappa}\nonumber\\
   (k_\nu)_{\lambda\kappa} &=& \frac{g}{2M}\,m_\nu^\lambda\,\delta_\lambda^\kappa\nonumber\\
  (k_e)_{\lambda\kappa} &=& \frac{g}{2M}\,m_e^\mu\,\cx_{\lambda\mu}\delta_\mu^\rho \cx^\dagger_{\rho\kappa}\nonumber
\end{eqnarray}
It might seem at first sight that one can simply use
\eqref{massrelations} to define the matrices $k_x$ but this
overlooks the fact that \eqref{rescaledyukawamasses} implies one
constraint:
\begin{equation}\label{massrelation1}
\Tr(k_\nu^*k_\nu+k_e^*k_e+3(k_u^*k_u+k_d^*k_d))=\,2\, g^2\,,
\end{equation}
using \eqref{coeffymterm} to replace $\frac{\pi^2}{f_0}$ by $ 2\,
g^2$. When expressed in the right hand side \ie the standard model
parameters this gives
\begin{equation}\label{massrelation2}
\sum_\lambda\,(m^\lambda_\nu)^2+(m^\lambda_e)^2+3\,(m^\lambda_u)^2+3\,
(m^\lambda_d)^2=\,8\,M^2
\end{equation}
where $M$ is the mass of the $W$ boson. Thus with the standard
notation (\cite{knecht}) for the Yukawa couplings, so that the
fermion masses are $m_f=\frac {1}{\sqrt 2} y_f\,v$, $v=\frac
{2M}{g}$ the relation reads
\begin{equation}\label{massrelation3}
\sum_\lambda\,(y^\lambda_\nu)^2+(y^\lambda_e)^2+3\,(y^\lambda_u)^2+3\,
(y^\lambda_d)^2=\,4\,g^2
\end{equation}
Neglecting the other Yukawa coupling except for the top quark, and
imposing the relation \eqref{massrelation3} at unification scale,
then running it downwards using the renormalization group one gets
the boundary value $\frac{2}{\sqrt 3} g\sim 0.597$ for $y_t$ at
unification scale which gives a Fermi scale value of the order of
 $y_0= \sim 1.102$ and a top quark
mass of the order of $\frac {1}{\sqrt 2} y_0\,v \sim \,173\,y_0$
GeV. This is fine since a large neglected tau neutrino Yukawa
coupling (allowed by the see-saw mechanism) similar to that of the
top quark, lowers the value at unification by a factor of $\sqrt
\frac 34$ which  has the effect of lowering the value of $y_0$ to
$y_0\sim 1.04 $. This yields an acceptable value for the top quark
mass (whose Yukawa coupling is $y_0\sim 1  $), given that we still
neglected all other smaller Yukawa couplings.

\smallskip
 The conversion table \ref{smtospec} shows that all the
mass parameters of the standard model now acquire geometric meaning
as components of the noncommutative metric as displayed in the right
column.

\smallskip

\begin{table}
\begin{center}

  \medskip

  \begin{tabular}{|c||c|c||c|}
  \hline
  Standard Model & notation & notation & Spectral Action\\
  \hline
  & & &\\
 Higgs Boson & $\varphi=(\frac{2M}{g}+ H-i\phi^{0},-i\sqrt{2}\phi^{+})$ & $\higgs=\frac{1}{\sqrt 2}\frac{\sqrt{a}}
 {g} (1+\psi)$& Inner metric$^{(0,1)}$ \\
 &&&\\
  \hline
  & & &\\
  Gauge bosons & $A_\mu,Z^0_\mu, W^\pm_\mu, g_\mu^a$ & $(B,W,V)$ & Inner metric$^{(1,0)}$ \\
  &&&\\
  \hline
  & & &\\
   Fermion masses&   $m_u,m_\nu$ &   $M_u=\delta_u,M_\nu=\delta_\nu$  &  Dirac$^{(0,1)}$ in $u,\nu$\\
    $u,\nu$&&&\\
  \hline
  & && \\
  CKM matrix & $C_\lambda^\kappa, m_d$ & $M_d=C\,\delta_d\,C^\dagger$ &  Dirac$^{(0,1)}$ in $d$\\
  Masses down &&&\\
  \hline
   & && \\
  Lepton mixing & $\cx_{\lambda\kappa}, m_e$ & $M_e=\cx\,\delta_e\,\cx^\dagger$ &  Dirac$^{(0,1)}$ in $e$\\
  Masses leptons $e$ &&&\\
  \hline
  & && \\
  Majorana & $M^R$ & $M_R$ & Dirac$^{(0,1)}$ in $\nu_R, \bar \nu_R$\\
   mass matrix&&&\\
  \hline
  & && \\
  Gauge couplings & $g_1=g\,{\rm tg}(w),g_2=g,g_3=g_s$ & $g_{3}^2= g_{2}^2
  = \frac{5}{ 3} \, g_{1}^2$ & Fixed at \\
   &&&unification\\
  \hline
  & && \\
  Higgs scattering & $\frac 18\,g^2\,\alpha_h, \alpha_h=\frac{m_h^2}{4M^2} $ &
  $\lambda_0  =  g^2\,\frac{b}{ a^2}$ & Fixed at \\
   parameter&&&unification\\
  \hline
  & && \\
  Tadpole constant & $\beta_h, (-\alpha_{h}\,M^2\,+\frac{\beta_h}{2})\,|\varphi|^2$ &
  $\mu_0^2=2\frac{f_2\Lambda^2}{f_0}  -\frac ea$ & $- \mu_0^2\, |\higgs|^2$\\
   &&&\\
  \hline
  & && \\
  Graviton & $g_{\mu\nu}$ & $\dirac_M$  & Dirac$^{(1,0)}$ \\
   &&&\\
  \hline
\end{tabular}
\bigskip
\bigskip
  \caption{Conversion from Spectral Action to Standard Model}\label{smtospec}

\end{center}
\end{table}

\bigskip

\section{Interpretation}

 It is not clear what the physics meaning is since
unlike in grand unified theories one is still lacking a
renormalizable theory that would take over above the unification
energy. But one can nevertheless hope that such a theory will be
discovered along the lines of QFT on noncommutative spaces, or even
that the fundamental theory has selected a preferred scale and  is a
fully unified theory at the operator theoretic level (\ie a kind of
spectral random matrix theory where the operator $D$ varies in the
symplectic ensemble corresponding to the commutation with
$i=\sqrt{-1}$ and $J$ that generate the quaternions) of which the
standard model coupled with gravity is just a manifestation when one
integrates the high energy modes \'a la Wilson. Then following
\cite{cc2} one can take the value of the Higgs quartic self-coupling
\eqref{higgsscat} as an indication at that same energy and
(\cite{knecht}) get a rough estimate (around $170$ GeV) for the
Higgs mass under the ``big desert" hypothesis. It is satisfactory
that the prediction for the Weinberg angle (the same as $\SU(5)$
GUT) is not too far off and that the mass relation gives a sensible
answer. But it is of course very likely that instead of the big
desert one will meet gradual refinements of the noncommutative
geometry $M\times F$ when climbing in energy to the  unification
scale.

\smallskip The naturalness problem is of course still there, but
interestingly the new terms involving $M_R$  provide room for
obtaining in the spectral action a term that mimics the nasty
quadratic divergence, whose coefficient changes sign under the
running of the remormalization group. This freedom holds provided
that the number of generations is $>1$. The quadratic coupling is
 $\mu_0^2 = \ 2\,\frac{f_2\,\Lambda^2}{ f_0}-\frac{e}{a}$. The
 presence of the new term $-\frac{e}{a}$ (which was absent in
 \cite{cc2}) allows for the possibility that the sign of
  this mass term is arbitrary provided there are at least
 two generations. We shall assume to discuss this point that the
 matrix $M_R$ is a multiple of a fixed matrix
$k_R$ \ie is of the form $M_R=x\,k_R$. The value of $x$ is fixed by
the equations of motion of the spectral action \ie by minimizing the
cosmological term. It gives
\begin{equation}\label{mrstarmr}
x^2=\frac{2\,f_2\,\Lambda^2\,\Tr(k_R^*k_R)}{f_0\,\Tr((k_R^*k_R)^2)}\,,\quad
M_R^*M_R=\frac{2\,f_2\,\Lambda^2}{f_0}\,\frac{k_R^*k_R\,\Tr(k_R^*k_R)}{\Tr((k_R^*k_R)^2)}
\end{equation}
  Using \eqref{mrstarmr} and \eqref{rescaledyukawamasses} one gets
  \begin{equation}\label{muzerosimple}
\mu_0^2=\,2\,\Lambda^2\,\frac{f_2}{f_0}\,(1-X)\,,\quad
X=\frac{\Tr(k_R^*k_R\,k_\nu^*k_\nu)\,\Tr(k_R^*k_R)}{\Tr(k_\nu^*k_\nu
+k_e^*k_e+3(k_u^*k_u+k_d^*k_d))\Tr((k_R^*k_R)^2)}
\end{equation}
In order to compare $X$ with $1$ we need to determine the range of
variation of the largest eigenvalue of
$\rho(k_R)=\,\frac{k_R^*k_R\,\Tr(k_R^*k_R)}{\Tr((k_R^*k_R)^2)}$ as a
function of the number $N$ of generations. One finds that this range
of variation, for $k_R\in M_N(\C)$, is the interval
$$
[1,\frac 12(1+\sqrt N)]
$$

\medskip
This suffices to show that provided the number $N$ of generations is
$>1$, there is room to get a small value of $\mu_0^2$. Note that a
similar discussion applies to the cosmological term $\gamma_0$ which
inherits a negative contribution from the presence of the $M_R$
term.

\section{Appendix: Real Structure and inner fluctuations}\label{appenreal}

We just briefly recall here the definition of spectral triple
$(\cA,\cH,D)$ and of real structure  \cite{Coreal}:

\begin{defn} \label{spectripdef} A spectral triple $({\mathcal
A},{\mathcal H},D)$ is given by an involutive unital algebra $\cA$
represented as operators in a Hilbert space $\cH$ and a self-adjoint
operator $D$ with compact resolvent such that all commutators
$[D,a]$ are bounded for $a\in \cA$.
\end{defn}

A spectral triple is {\em even} if the Hilbert space $\cH$ is
endowed with a $\Z/2$- grading $\gamma$ which commutes with any
$a\in \cA$ and anticommutes with $D$.

\smallskip
\begin{defn}\label{realstr}
A real structure of $KO$-dimension  $n\in \Z/8$ on a spectral triple
$(\cA,\cH,D)$ is an antilinear isometry $J: \cH \to \cH$, with the
property that
\begin{equation}\label{per8}
J^2 = \varepsilon, \ \ \ \ JD = \varepsilon' DJ, \ \ \text{and} \ \
J\gamma = \varepsilon'' \gamma J \, \text{(even case)}.
\end{equation}
The numbers $\varepsilon ,\varepsilon' ,\varepsilon'' \in \{ -1,1\}$
are a function of $n \mod 8$ given by

\begin{center}
\begin{tabular}
{|c| r r r r r r r r|} \hline {\bf n }&0 &1 &2 &3 &4 &5 &6 &7 \\
\hline \hline
$\varepsilon$  &1 & 1&-1&-1&-1&-1& 1&1 \\
$\varepsilon'$ &1 &-1&1 &1 &1 &-1& 1&1 \\
$\varepsilon''$&1 &{}&-1&{}&1 &{}&-1&{} \\  \hline
\end{tabular}
\end{center}

Moreover, the action of $\cA$ satisfies the commutation rule
\begin{equation}\label{comm-rule}
[a,b^0] = 0 \quad \forall \, a,b \in \cA,
\end{equation}
where
\begin{equation}\label{b0}
b^0 = J b^* J^{-1} \qquad \forall b \in \cA,
\end{equation}
and the operator $D$ satisfies
\begin{equation}\label{order1}
[[D,a],b^0] = 0 \qquad \forall \, a,b \in \cA \, .
\end{equation}
\end{defn}

\medskip
The key role of the real structure $J$ is to yield the following
{\em adjoint action} of the unitary group $\cU$ of the algebra $\cA$
on the hilbert space $\cH$ (of spinors). One defines a right
$\cA$-module structure on $\cH$ by
\begin{equation}\label{bimodule}
\xi\,b=\,b^0\,\xi\qqq \;\xi \in \cH\,,\quad b\in \cA
\end{equation}
The unitary group of the algebra $\cA$ then acts by the ``adjoint
representation" in $\cH$ in the form
\begin{equation}\label{adjact}
\xi \in \cH\to {\rm Ad}(u)\,\xi=u\,\xi\,u^*\qqq \;\xi \in
\cH\,,\quad u\in \cA \,,\quad u\,u^*=u^*\,u=1\,,
\end{equation}
and the inner fluctuation of the metric is given by
\begin{equation}\label{innerfluc1} D\to D_A=D+A+\varepsilon'
\,J\,A\,J^{-1}
\end{equation}
where  $A$ is a self-adjoint operator of the form
\begin{equation}\label{innerfluc2}
A=\sum\,a_j[D,b_j]\,,\quad a_j,b_j\in \cA.
\end{equation}
The {\em unimodular} inner fluctuations are obtained by restricting
to those $A$ which are traceless \ie fulfill the condition
$\Tr(A)=0$.

\section{ Aknowledgements}

The detailed computations and extension of this work to the
left-right model  will appear in a joint work with Ali Chamseddine
and Matilde Marcolli. The need to have independence between the
$KO$-dimension and the metric dimension already emerged implicitly
in the work of L. D\c{a}browski and  A. Sitarz  on Podle\'s quantum
spheres \cite{dab}. The results of this work were announced in a
talk at the Newton Institute in July 2006, and the fear of a
numerical error in the above computations delayed the present
publication. It is a pleasure to acknowledge the independent
preprint by John Barrett (A Lorentzian version of the
non-commutative geometry of the standard model of particle physics)
with a similar solution of the fermion doubling problem which
accelerated the present publication.


\begin{thebibliography}{99}




 \bibitem{cc1}  A.~Chamseddine, A.~Connes, {\em  Universal Formula for Noncommutative Geometry Actions: Unification of
Gravity and the Standard Model}, Phys. Rev. Lett. 77, 486804871
(1996).

\bibitem{cc2} A.~Chamseddine, A.~Connes, {\em  The Spectral Action Principle}, Comm. Math. Phys. 186, 731-750 (1997).


\bibitem{cc3} A.~Chamseddine, A.~Connes, {\em  Scale Invariance in the Spectral Action}, hep-th/0512169 to appear in Jour.
 Math. Phys

\bibitem{cc4} A.~Chamseddine, A.~Connes, {\em Inner fluctuations of the spectral
action}, hep-th/0605011.

\bibitem{Coleman} S.~Coleman, {\em Aspects of symmetry}, Selected Erice
Lectures, Cambridge University Press, 1985.



\bibitem{Co-book}  A.~Connes, {\it Noncommutative geometry},
 Academic Press (1994).


 \bibitem{Coreal} A.~Connes, {\em  Non commutative geometry and
reality},   Journal of Math. Physics  36 no. 11 (1995).



 \bibitem{CoSM} A.~Connes, {\em Gravity coupled with matter and the
foundation of noncommutative geometry}, Comm. Math. Phys. (1995)

 \bibitem{CoMM} A.~Connes, M.~Marcolli {\em Quantum fileds, noncommutative spaces and motives},
Book in preparation.


\bibitem{dab}  L.~D\c{a}browski, A.~Sitarz, {\em
Dirac operator on the standard Podle\'s quantum sphere},
Noncommutative Geometry and Quantum Groups, Banach Centre
Publications 61, Hajac, P.~M. and Pusz, W. (eds.), Warszawa: IMPAN,
2003, pp. 49--58.



\bibitem{gbis} J.~Gracia-Bondía, B.~Iochum, T.~Schucker, {\em The standard model in noncommutative
geometry and fermion doubling}. Phys. Lett. B 416 no. 1-2 (1998),
123--128.


\bibitem {kastler} D.~Kastler, {\em  Noncommutative geometry and fundamental
physical interactions: The Lagrangian level}, Journal. Math. Phys.
41 (2000), 3867-3891.

\bibitem{knecht} M.~Knecht, T.~Schucker {\em Spectral action and big
desert} hep-ph/065166

\bibitem{reuter} O.~Lauscher, M.~Reuter,
{\em  Asymptotic Safety in Quantum Einstein Gravity: nonperturbative
renormalizability and fractal spacetime structure}, hep-th/0511260



\bibitem{lizzi} F.~Lizzi, G.~Mangano, G.~Miele, G.~Sparano, {\em
Fermion Hilbert space and Fermion Doubling in the Noncommutative
Geometry Approach to Gauge Theories} hep-th/9610035.

 \bibitem{MoPa} R.N.~Mohapatra, P.B.~Pal, {\em Massive neutrinos in
physics and astrophysics}, World Scientific, 2004.

\bibitem{VDiag} M.~Veltman, {\em Diagrammatica: the path to Feynman
diagrams}, Cambridge Univ. Press, 1994.



\end{thebibliography}
\end{document}